%% file: paper.tex
\documentclass[10pt, conference]{IEEEtran}
\ifCLASSINFOpdf
\else
\fi
\usepackage{cite}
\usepackage{amsmath,amssymb,amsfonts}
\usepackage{algorithmic}
\usepackage{graphicx}
\usepackage{textcomp}
\usepackage{xcolor}
\def\BibTeX{{\rm B\kern-.05em{\sc i\kern-.025em b}\kern-.08em
    T\kern-.1667em\lower.7ex\hbox{E}\kern-.125emX}}
    
\usepackage{color,xcolor}
\usepackage{multirow}
\usepackage{booktabs} 
\usepackage{comment}
\usepackage{tablefootnote}
\usepackage{ifthen}
\usepackage{url}
\usepackage{amsmath}
\usepackage{threeparttable}
\usepackage{hyphenat}
\usepackage{bbding}
\usepackage{xspace}
\usepackage[T1]{fontenc}
\usepackage[ruled,linesnumbered]{algorithm2e}
\usepackage{array,multirow,graphicx}
\usepackage{float}
\usepackage{balance}
\usepackage{hyperref}
\usepackage{tikz}
\usepackage{calc}
\usepackage{subfigure}
\usepackage{listings,amsfonts}
\usepackage{pifont}
\usepackage{url}
\usepackage{balance}
\usepackage{xspace}
\usepackage{hyperref,epsfig,endnotes}
\usepackage{array}
\usepackage{paralist}
\usepackage{multirow,makecell}
\usepackage[normalem]{ulem}
\usepackage{pgfplots}
\usepackage{tcolorbox}
\usepackage{courier}
\usepackage{enumitem}
\usepackage{soul}
\usepackage{comment}

\begin{document}
%
\title{Benchmarking Software Vulnerability Detection Techniques: A Survey}

\author{
\IEEEauthorblockN{Yingzhou Bi}
\IEEEauthorblockA{
Nanning Normal University\\
Nanning, China\\
byz@nnnu.edu.cn
}

\and
\IEEEauthorblockN{Jiangtao Huang}
\IEEEauthorblockA{
Nanning Normal University\\
Nanning, China}

\and
\IEEEauthorblockN{Peihui Liu}
\IEEEauthorblockA{
Nanning Normal University\\
Nanning, China}

\and
\IEEEauthorblockN{Lianmei Wang}
\IEEEauthorblockA{
Nanning Normal University\\
Nanning, China}

}


%


\maketitle

\begin{abstract}
Software vulnerabilities can have serious consequences, which is why many techniques have been proposed to defend against them. Among these, vulnerability detection techniques are a major area of focus. However, there is a lack of a comprehensive approach for benchmarking these proposed techniques. In this paper, we present the first survey that comprehensively investigates and summarizes the current state of software vulnerability detection benchmarking.
We review the current literature on benchmarking vulnerability detection, including benchmarking approaches in technique-proposing papers and empirical studies. We also separately discuss the benchmarking approaches for traditional and deep learning-based vulnerability detection techniques. Our survey analyzes the challenges of benchmarking software vulnerability detection techniques and the difficulties involved.
We summarize the challenges of benchmarking software vulnerability detection techniques and describe possible solutions for addressing these challenges.
\end{abstract}


%
\IEEEpeerreviewmaketitle

\input{intro.tex}
\input{techniques.tex}

\input{empirical.tex}
\input{challenges.tex}

\input{conclusion.tex}

\bibliography{paper}
\bibliographystyle{IEEEtran}




\end{document}

%% file: intro.tex
\section{Introduction}
Software vulnerabilities can have a significant impact and result in serious consequences for modern software. They cause billions of dollars in losses each year in large software systems~\cite{realvul2020}. As a result, effective and efficient detection and repair of software vulnerabilities are crucial. To defend against vulnerabilities, many techniques have been proposed. Among them, vulnerability detection is one of the most important domains of defense. Consequently, numerous vulnerability detection techniques have been proposed and have achieved success in their evaluations.


However, there is a lack of understanding and benchmarking of vulnerability detection techniques. Some papers only propose techniques without evaluating them with benchmarking datasets~\cite{holzmann2002static, kroening2014cbmc, cuoq2012frama}. Other papers simply evaluate the techniques by computing the number of vulnerabilities found in several real-world projects~\cite{austin2011one, austin2013comparison, pozza2006comparing, antunes2009comparing}. However, their benchmarking datasets do not have ground truths, making it impossible to compute precision, recall, and F1 scores. As a result, it is challenging to benchmark the techniques fairly and comprehensively.


In this paper, we review the current literature on benchmarking vulnerability detection techniques. First, we examine the benchmarking approaches used when introducing these techniques. Next, we review third-party empirical studies on benchmarking the techniques. Finally, we summarize the challenges encountered during benchmarking and provide an overview of possible solutions for addressing these challenges.


In summary, we make the following contributions:
\begin{itemize}
    \item We analyze the lack of benchmarking for software vulnerability detection and the difficulties that arise from this lack.
    \item To the best of our knowledge, our study is the first to comprehensively investigate and summarize the current state of software vulnerability detection benchmarking.
    \item Based on our survey, we discuss the major challenges associated with benchmarking software vulnerability detection and describe some possible solutions for addressing these challenges in the future.
\end{itemize}

%% file: techniques.tex
\section{Existing Benchmarking Approaches}
Over the past decade, numerous vulnerability detection techniques and empirical studies have been proposed. Correspondingly, many vulnerability detection benchmarking approaches have been used to evaluate these techniques. In this section, we review the existing benchmarking approaches used to evaluate these techniques.
We first examine the benchmarking approaches used in the papers that propose vulnerability detection techniques. Next, we review the approaches used in empirical studies that compare and evaluate existing vulnerability detection techniques to understand their strengths and weaknesses.

\subsection{Technique Proposing Papers}
It is common for technical papers to use benchmarking approaches to evaluate the techniques they introduce. However, during our literature study, we noticed that the benchmarking approaches used before and after the trend of DL-based vulnerability detection are significantly different.

Specifically, traditional vulnerability detection techniques (such as static analysis, dynamic analysis, and penetration analysis) before the emergence of DL-based techniques were typically evaluated through technical reviews and case studies. In contrast, DL-based techniques are often evaluated on large-scale vulnerability datasets due to the need for numerous data to train and test them. Therefore, we separately discuss the benchmarking approaches used in papers on traditional vulnerability detection techniques and those on DL-based techniques.


\subsubsection{Traditional Vulnerability Detection Techniques}
Since traditional vulnerability detection techniques are mostly deterministic, their technical strengths and weaknesses can be relatively easy to analyze by humans. Thus, some traditional vulnerability detection techniques evaluate themselves by reviewing their technical strengths and weaknesses only:


UNO~\cite{holzmann2002static} is a static analyzer that detects software vulnerabilities by checking user-defined properties. It evaluates itself by reviewing its features and side effects, such as in which scenarios UNO can detect vulnerabilities. CBMC~\cite{kroening2014cbmc} is a static bounded model checker for vulnerability detection, which evaluates itself by discussing its strengths and weaknesses. Frama-C~\cite{kirchner2015frama, cuoq2012frama} is a C program verification tool for vulnerability detection, which evaluates itself by discussing its technical features. ITS4~\cite{viega2000its4} is a vulnerability scanner for C and C++ code, which also evaluates itself by reviewing its features, strengths, and weaknesses, while comparing it with other techniques and describing practical experiences of using it. Zhu's model~\cite{zhu2022model} is evaluated based on which vulnerability-relevant properties it can check. ESC/Java~\cite{flanagan2002extended} is an extended vulnerability checker for Java, which evaluates itself by discussing the user experience in different scenarios. These techniques' benchmarking approaches are based on feature review and user experience, without applying them to vulnerability detection test cases, which makes it difficult to compare them with each other.


In comparison, some other techniques evaluate themselves by conducting case studies to better convince users. Undangle~\cite{caballero2012undangle} uses case studies on Firefox to demonstrate its ability to detect dangling pointers early. Valgrind MemCheck~\cite{nethercote2003valgrind, nethercote2007valgrind, seward2005using} applies itself to Open Office~\cite{ven2006introduction} to showcase its capability in finding real-world vulnerabilities. FindBugs~\cite{hovemeyer2004finding} collects user experiences from applying it to a geographic information system and a financial application to demonstrate its effectiveness. FlowDist~\cite{fu2021flowdist} evaluates recall and precision through manual evaluation on selected cases. Astree~\cite{blanchet2003static} applies itself to a project with 132,000 lines of code to demonstrate its effectiveness and efficiency in vulnerability detection. CodeSonar~\cite{jetley2008static} is evaluated on medical device software to discuss its strengths and weaknesses. AddressSanitizer~\cite{serebryany2012addresssanitizer} is applied to Google Chrome to demonstrate its ability to find real-world vulnerabilities. While these case studies apply the techniques to real-world vulnerability detection scenarios, the cases used are not large-scale, and different techniques use different cases, making it difficult to benchmark and compare techniques comprehensively and fairly.


Other papers have used large-scale standard benchmarking approaches to evaluate the effectiveness of vulnerability detection techniques. For instance, Facebook Infer~\cite{harmim2019scalable} used the benchmarking dataset introduced in CPROVER~\cite{kroening2016sound} to evaluate the performance of detecting deadlocks. MemSafe~\cite{simpson2013memsafe} applied itself to BugBench suite~\cite{lu2005bugbench} to evaluate its ability to detect memory safety errors. KLEE~\cite{cadar2008klee} was evaluated on multiple real-world test suites to test its error coverage. PCA~\cite{li2020pca}, Valgrind~\cite{nethercote2003valgrind}, DrMemory~\cite{bruening2011practical}, MemorySanitizer~\cite{stepanov2015memorysanitizer}, and AddressSanitizer~\cite{serebryany2012addresssanitizer} were evaluated on SPEC CPU2000~\cite{henning2000spec} and SPEC CPU2006~\cite{henning2006spec} to test the running time and memory usage of the vulnerability detection on real-world projects. However, these papers were published before the trend of DL-based vulnerability detection. Additionally, some benchmarking datasets, such as SPEC CPU2006, were not originally designed for vulnerability detection benchmarking but for other proposes (e.g., testing running time and memory usage). Thus, they can only be used to evaluate the efficiency of the techniques or check the number of vulnerabilities that can be found. Because the benchmarking datasets lack vulnerability ground truths, important evaluation metrics such as recall, precision, and F1 score cannot be evaluated.


A few papers have used vulnerability benchmarking datasets with ground truths to evaluate their techniques comprehensively. PolyCruise~\cite{li2022polycruise} is a dynamic flow analysis technique for multilingual software vulnerability detection. To enable a comprehensive evaluation, the authors of PolyCruise manually built a benchmarking dataset called PyCBench~\cite{li2022polycruise}, which contains ground truths to evaluate the technique. BOFSanitizer~\cite{wang2021bofsanitizer} is evaluated for detecting buffer overflow vulnerabilities using the Juliet Test Suite~\cite{black2018juliet}. VulSlicer~\cite{salimi2022vulslicer}, MVP~\cite{xiao2020mvp}, and LEOPARD~\cite{du2019leopard} are evaluated on the real-world vulnerability database CVE/NVD~\cite{booth2013national} for vulnerability detection. These evaluations against vulnerability benchmarking datasets enable crucial metrics such as recall, precision, and F1, which are essential for benchmarking the techniques fairly and comprehensively.


\subsubsection{DL-based Vulnerability Detection Techniques}
In recent years, there has been an increasing use of deep learning (DL) for software vulnerability detection. An empirical study on DL-based vulnerability detection techniques reveals that more than 55 techniques were proposed between 2016 and 2020~\cite{nong2022open}. Because DL-based models require a large amount of data to train and test, DL-based software vulnerability detection techniques use various benchmarking approaches to evaluate them. These benchmarking approaches cover different dimensions: training/testing data, granularity, and real-world case studies. We will discuss each of these dimensions separately below.


\textbf{Training/Testing Data: }
In 2018, Li et al.~\cite{li2018vuldeepecker} proposed VulDeePecker, which was the first DL-based vulnerability detector that achieved great success. The authors used 10,691 vulnerability samples to train and test the model. These samples had ground truths that indicated whether the given sample was vulnerable or non-vulnerable. This enabled the evaluation metrics such as accuracy, recall, precision, and F1 score. Among them, 9,851 samples were from SARD~\cite{black2017sard}, which were synthetic samples (i.e., artificially generated without considering the realism of the samples), and 840 samples were from CVE/NVD~\cite{nvd}, which were from real-world software projects. Subsequently, many papers used the same benchmarking approach and evaluation dataset (although the number of samples may differ slightly, they were from SARD and CVE/NVD) to evaluate their proposed new techniques~\cite{li2021sysevr, zou2019mu, zagane2020deep, wu2022vulcnn, tang2022sevuldet, thapa2022transformer,cao2022mvd,pinconschitenet}.

However, most vulnerability samples used for DL-based techniques are synthetic, making them too simple and easy for DL models to learn, leading to inflated results~\cite{chakraborty2021deep}. Therefore, using real-world vulnerability samples is more promising. Some techniques, like Draper~\cite{russell2018automated}, collected 1.27 million samples and used static analyzers to label vulnerable code to build their benchmarking dataset. However, the static analyzers' low accuracy, precision, recall, and F1 on real-world samples~\cite{nong2020preliminary, nong2021evaluating} make training and testing datasets very noisy. To address this, Devign~\cite{zhou2019devign} manually collected 22,361 vulnerability samples from 2 real-world projects, taking 600 man-hours. The benchmarking dataset has relatively balanced vulnerable and non-vulnerable samples, but in real-world projects, vulnerable code is the minority, making a balanced dataset unrealistic~\cite{nong2022open}.


Several DL-based vulnerability detection techniques use real-world samples without balancing to build benchmarking datasets for evaluation. For instance, TRL~\cite{lin2018cross} and RLMD~\cite{lin2019software} use 39,942 vulnerability samples from six real-world projects as their benchmarking datasets, where only 1.46\% of the samples are vulnerable. ReVeal~\cite{chakraborty2021deep} manually collects 18,169 vulnerability samples, of which only 9.16\% are vulnerable. On the other hand, IVDetect~\cite{li2021vulnerability}, LineVD~\cite{hin2022linevd}, muVulPreter~\cite{zou2022mvulpreter}, LineVul~\cite{fu2022linevul}, and Cheng~\cite{cheng2022path} use Big-Vul~\cite{fan2020ac}, which contains more than 168,000 non-vulnerable samples and 10,000 vulnerable samples. These benchmarking datasets can better represent real-world vulnerability detection scenarios.


\textbf{Granularity:} 
In addition to the datasets used for training and testing DL models, granularity in vulnerability detection is also crucial. In real-world vulnerability detection scenarios, developers not only need to know if a file or function is vulnerable, but they also need to identify the trigger point of the vulnerability (e.g., one or more statements that can be exploited) and the type of vulnerabilities. Traditional vulnerability detection techniques can easily provide detailed information because they analyze the code by examining whether statements contain potential trigger points for attacks. However, DL-based vulnerability detection models are black boxes, and we can only train them based on the task formulation and the training data.


Based on our literature study, most of the DL-based vulnerability detection techniques use function-level granularity (i.e., the models decide whether a given function is vulnerable or not), without considering the vulnerability types~\cite{zagane2020deep, wu2022vulcnn, tang2022sevuldet, thapa2022transformer,cao2022mvd, chakraborty2021deep,zhou2019devign,lin2018cross,lin2019software}. VulDeePecker~\cite{li2018vuldeepecker}, SySeVR~\cite{li2021sysevr}, uVulDeePecker~\cite{zou2019mu}, and VulDeeLocator~\cite{li2020vuldeelocator} consider code slice-level vulnerability detection, which is more fine-grained and easier for developers to understand and fix the vulnerabilities. uVulDeePecker~\cite{zou2019mu} not only achieves code slice-level detection, but also is able to classify the vulnerability types. IVDetect~\cite{li2021vulnerability}, LineVD~\cite{hin2022linevd}, mVulPreter~\cite{zou2022mvulpreter}, and LineVul~\cite{fu2022linevul} even consider statement-level vulnerability detection~\cite{li2021sysevr}. Based on the chronological order of these papers, detecting vulnerabilities in fine-grain granularities (e.g., statement-level) and providing the vulnerability types should be the future trend, because that is the need of real-world developers.

\textbf{Real-world Case Studies:} Ideally, the DL-based vulnerability detection techniques should not only show the high accuracy on the used benchmarking datasets, but also should be usable for real-world vulnerability detection. Thus, to show the usability of the techniques, some of the papers apply their techniques to real-world software projects and successfully find unknown vulnerabilities. Devign~\cite{zhou2019devign} selects the latest 112 vulnerability samples to check whether it has the capability to detect zero-day vulnerabilities. VulDeePecker~\cite{li2018vuldeepecker}, SySeVR~\cite{li2021sysevr}, VulCNN~\cite{wu2022vulcnn}, VulDeeLocator~\cite{li2020vuldeelocator}, and VulHunter~\cite{guo2019vulhunter} apply their techniques to real-world software projects and successfully find unknown vulnerabilities. The case studies on real-world scenarios significantly show the practicability of the techniques and should be an important benchmarking approach for vulnerability detection techniques.

%% file: empirical.tex
\subsection{Empirical Study Papers}
Besides the papers that introduce vulnerability detection techniques, there are also many empirical studies that aim to fairly and comprehensively evaluate the existing vulnerability detection techniques, so that they can find the weaknesses and strengths of the techniques and provide practical insights and suggestions for future technique development. To do that, these empirical studies also need to find effective benchmarking approaches to evaluate the techniques. Since the empirical studies need to cover many different vulnerability detection techniques, the benchmarking approaches need to be general, fair, and standalone, which have higher standards than the benchmarking approaches in the technique proposing papers. 

In this section, we review the existing empirical studies on vulnerability detection techniques, and summarize the better benchmarking approaches for traditional and DL-based vulnerability detection techniques.

\subsubsection{Traditional Vulnerability Detection Techniques}
Many empirical studies for benchmarking vulnerability detection techniques exist and they use different approaches to evaluate and compare them. Some studies compare and evaluate different static code analysis approaches against SQL injection and XSS attack vulnerabilities in web services`\cite{antunes2009comparing,5552783,fonseca2007testing}. Some other studies evaluate buffer overflow vulnerability detectors. However, they only target specific vulnerability types and domains which limit the comprehensiveness of the studies. Also, these studies benchmark the evaluated techniques using a relatively small number of samples, thus many cases are missed by the techniques that are not actually evaluated.

Some other studies use a relatively large number of samples to benchmark vulnerability detection techniques~\cite{kang2022detecting}. Austin et al.~\cite{austin2011one, austin2013comparison} uses three electronic health record systems to benchmark automated penetration testing and static analysis techniques. In some other studies~\cite{pozza2006comparing, antunes2009comparing}, the authors evaluate the vulnerability detection techniques by counting the number of vulnerabilities found in synthetic or real-world software projects. However, these evaluations do not refer to any ground truth. Thus, they cannot compute precision, recall, and F1 score which are important metrics in their studies. 

From the cases above, we can see that the benchmarking metrics are important for the empirical studies on vulnerability detection techniques. Thus, in ~\cite{5552783}, the authors discuss and provide the metrics for benchmarking vulnerability detection techniques, but the actual empirical experiments on the state-of-the-art techniques are not performed. Another technical report~\cite{shirey2000internet} simply discusses the capabilities of the evaluated vulnerability detection techniques without any analytical experiments, thus no empirical comparisons are conducted, making the actual detection performance not able to be assessed. 

A few empirical studies~\cite{antunes2010benchmarking, fonseca2007testing} evaluate and compare vulnerability detection techniques using samples with ground truths, thus they can compute precision, recall, and F1. However, these studies only evaluate commercial tools, thus the reasons behind the successes and failures are difficult to assess.

In comparison, Yu et al. conduct empirical studies on five open-source vulnerability detection techniques against datasets with ground truths~\cite{nong2020preliminary, nong2021evaluating}. The datasets cover 20 categories of memory related vulnerabilities. Thus, the studies benchmark general vulnerability detection techniques and the precision, recall, F1 score, as well as the efficiency are comprehensively evaluated. They also do several in-depth case studies to analyze the reasons behind the successes and failures. However, in the datasets they used to evaluate, most of the samples are synthetic, making the evaluation not very referable to real-world vulnerability detection scenarios.

\subsubsection{DL-based Vulnerability Detection Techniques}
Because of the trend of DL-based vulnerability detection techniques, many empirical studies are also proposed on these DL-based vulnerability detection techniques. Li et al.~\cite{li2019comparative} did a comparative study of deep learning-based vulnerability detection tools. They quantitatively evaluate different factors' impact on the vulnerability detection performance. However, they only simply use several different deep learning models without vulnerability detection specification to do the evaluation.
Fernandez et al.~\cite{fernandez2019case} did a case study for network intruction detection using deep learning models. Jabeen et al.~\cite{jabeen2022machine} did a comparative study on machine learning techniques for software prediction. However, the above techniques only evaluate the DL models without using large-scale real-world vulnerability datasets, making their evaluation not comprehensive. Zhang~\cite{zhang2023survey} did a survey on the learning-based vulnerability repair techniques.

In comparison, Mazuera-Rozo et al.~\cite{mazuera2021shallow} uses different source code representations to compare the effectiveness of deep learning models. They use both real-world, highly imbalanced dataset and a balanced dataset for comparison. Steenhoek et al.~\cite{steenhoek2022empirical} did an empirical study of deep learning models for vulnerability detection. They investigate six research questions on nine state-of-the-art deep learning models against two widely used vulnerability detection datasets. Siow~\cite{siow2022learning} did an empirical study on program semantics learning with different code representations. They evaluate different mainstream code representations including feature-based, sequence-based, tree-based, and graph-based. 

However, most of the above empirical studies only evaluate several different deep learning models or source code representation for deep learning models. It is difficult for them to find available and replicable tools to do their comparative studies. Thus, Nong et al.~\cite{nong2022open} conduct a comprehensive empirical study of the open science status on deep learning-based vulnerability detection. They evaluate the availability, excitability, reproducibility, and replicability of 55 existing vulnerability detection techniques. Their in-depth study investigates the source code links in the papers, tool documentation and completeness, tool reproduction which uses the original datasets, and tool replication which uses third-party real-world datasets. 

%% file: challenges.tex
\section{Challenges and Solutions}
Based on the survey above, we notice that there are several challenges for benchmarking vulnerability detection:

\begin{itemize}
    \item First, while many of the papers propose vulnerability detection techniques or build their evaluation datasets, their source code and datasets are not fully released to other researchers, making the benchmarking from other researchers difficult.
    \item Second, there are many different kinds of evaluation metrics. Some evaluations use the numbers of vulnerabilities the techniques can find to evaluate. Some other evaluations use samples with ground truths to compute recall, precision, and F1 score. Besides, there are many different factors that can impact the evaluation, such as the realism of the evaluation datasets, the data balance of the datasets, the granularity of the samples (e.g., line, function, file or project), training dataset size for deep learning models, and the complexity of the samples (e.g., code structure and control flow complexity). These should be also comprehensively considered for the benchmarking.
    \item Third, there is a lack of large-scale and real-world vulnerability datasets for benchmarking vulnerability detection. Many of them use small (only several hundred samples) datasets to evaluate the vulnerability detection techniques. This may be enough for traditional vulnerability detection techniques, but not enough for deep learning-based vulnerability detection because they need many samples (usually >10,000) to train the models well.

\end{itemize}

To solve the three challenges above, some solutions have been tried for solving the challenges. Nong et al.~\cite{nong2022open} evaluate the availability, excitability, reproducibility, and replicability of existing vulnerability detection techniques and provide actionable suggestions to solve the first challenge. Chakraborty et al.~\cite{chakraborty2021deep} build their own dataset and evaluate the impact of data realism and data balance on the effectiveness, which help solve the second challenge. Croft et al.~\cite{croft2023data} and liu et al.~\cite{liu2022investigating} evaluate the data quality of the vulnerability datasets, which help solve the third challenge. Indeed, there are many independent works trying to build high-quality vulnerability datasets. Some of them collect the respective source code from the publicly available CVE/NVD databases~\cite{bhandari2021cvefixes, fan2020ac, reis2021ground}. Some of them propose techniques that automatically collect vulnerability data in the wild~\cite{xu2021tracer, sawadogo2021early, sawadogo2022sspcatcher, nguyen2022hermes, nguyen2022vulcurator, guo2022hyvuldect, zheng2021d2a, woo2021v0finder,li2022polyfax}. 

However, these datasets are still not large-scale for training effective deep learning-based vulnerability detection techniques. Recently, there are new approaches proposed which automatically generate high-quality and large-scale vulnerability dataset. Nong et al.~\cite{nong2022generating} try to use existing deep learning-based code repair/edit tools to inject vulnerability into existing real-world normal programs and show the feasibility of doing so. He et al.~\cite{he2023controlling} develop a tool which generates both secure and vulnerable code by controlling large language models.
Nong et al. did similar work~\cite{nong2022vulgen} which combines the advantages of traditional pattern application and deep learning-based vulnerability injection localization to generate more high-quality vulnerability datasets. This indicates that generating vulnerability data may be a good way to build high-quality and large-scale datasets.

%% file: conclusion.tex
\section{Conclusion}
In this paper, we did the first survey that comprehensively investigates and summarizes the status of software vulnerability detection benchmarking. We review the current literature on benchmarking vulnerability detection, including the benchmarking approaches in technique proposing papers, empirical studies. We also separately discuss the benchmarking approaches for traditional and deep learning-based vulnerability detection techniques. Our survey analyzes the lack of software vulnerability detection benchmarking and the respective difficulties. 
We summarize the challenges of benchmarking software vulnerability detection techniques and describe the possible solutions for solving the challenges.

%% file: paper.bbl
\begin{thebibliography}{10}
\providecommand{\url}[1]{#1}
\csname url@samestyle\endcsname
\providecommand{\newblock}{\relax}
\providecommand{\bibinfo}[2]{#2}
\providecommand{\BIBentrySTDinterwordspacing}{\spaceskip=0pt\relax}
\providecommand{\BIBentryALTinterwordstretchfactor}{4}
\providecommand{\BIBentryALTinterwordspacing}{\spaceskip=\fontdimen2\font plus
\BIBentryALTinterwordstretchfactor\fontdimen3\font minus
  \fontdimen4\font\relax}
\providecommand{\BIBforeignlanguage}[2]{{%
\expandafter\ifx\csname l@#1\endcsname\relax
\typeout{** WARNING: IEEEtran.bst: No hyphenation pattern has been}%
\typeout{** loaded for the language `#1'. Using the pattern for}%
\typeout{** the default language instead.}%
\else
\language=\csname l@#1\endcsname
\fi
#2}}
\providecommand{\BIBdecl}{\relax}
\BIBdecl

\bibitem{realvul2020}
F.~Civaner, ``Real-life software security vulnerabilities and what you can do
  to stay safe,''
  https://hackernoon.com/how-software-security-vulnerabilities-work-and-what-you-can-do-to-stay-safe-c9596d993581.

\bibitem{holzmann2002static}
G.~Holzmann, ``Static source code checking for user-defined properties,'' in
  \emph{Proc. IDPT}, vol.~2, 2002.

\bibitem{kroening2014cbmc}
D.~Kroening and M.~Tautschnig, ``Cbmc--c bounded model checker,'' in
  \emph{International Conference on Tools and Algorithms for the Construction
  and Analysis of Systems}.\hskip 1em plus 0.5em minus 0.4em\relax Springer,
  2014, pp. 389--391.

\bibitem{cuoq2012frama}
P.~Cuoq, F.~Kirchner, N.~Kosmatov, V.~Prevosto, J.~Signoles, and B.~Yakobowski,
  ``Frama-c,'' in \emph{International conference on software engineering and
  formal methods}.\hskip 1em plus 0.5em minus 0.4em\relax Springer, 2012, pp.
  233--247.

\bibitem{austin2011one}
A.~Austin and L.~Williams, ``One technique is not enough: A comparison of
  vulnerability discovery techniques,'' in \emph{International Symposium on
  Empirical Software Engineering and Measurement}.\hskip 1em plus 0.5em minus
  0.4em\relax IEEE, 2011, pp. 97--106.

\bibitem{austin2013comparison}
A.~Austin, C.~Holmgreen, and L.~Williams, ``A comparison of the efficiency and
  effectiveness of vulnerability discovery techniques,'' \emph{Information and
  Software Technology}, vol.~55, no.~7, pp. 1279--1288, 2013.

\bibitem{pozza2006comparing}
D.~Pozza, R.~Sisto, L.~Durante, and A.~Valenzano, ``Comparing lexical analysis
  tools for buffer overflow detection in network software,'' in \emph{1st
  International Conference on Communication Systems Software \&
  Middleware}.\hskip 1em plus 0.5em minus 0.4em\relax IEEE, 2006, pp. 1--7.

\bibitem{antunes2009comparing}
N.~Antunes and M.~Vieira, ``Comparing the effectiveness of penetration testing
  and static code analysis on the detection of {SQL} injection vulnerabilities
  in web services,'' in \emph{Pacific Rim International Symposium on Dependable
  Computing}, 2009, pp. 301--306.

\bibitem{kirchner2015frama}
F.~Kirchner, N.~Kosmatov, V.~Prevosto, J.~Signoles, and B.~Yakobowski,
  ``Frama-c: A software analysis perspective,'' \emph{Formal Aspects of
  Computing}, vol.~27, no.~3, pp. 573--609, 2015.

\bibitem{viega2000its4}
J.~Viega, J.-T. Bloch, Y.~Kohno, and G.~McGraw, ``Its4: A static vulnerability
  scanner for c and c++ code,'' in \emph{Proceedings 16th Annual Computer
  Security Applications Conference (ACSAC'00)}.\hskip 1em plus 0.5em minus
  0.4em\relax IEEE, 2000, pp. 257--267.

\bibitem{zhu2022model}
W.~Zhu, ``Model checking for alphacode-generated programs,'' in \emph{2022 7th
  International Conference on Intelligent Computing and Signal Processing
  (ICSP)}.\hskip 1em plus 0.5em minus 0.4em\relax IEEE, 2022, pp. 794--798.

\bibitem{flanagan2002extended}
C.~Flanagan, K.~R.~M. Leino, M.~Lillibridge, G.~Nelson, J.~B. Saxe, and
  R.~Stata, ``Extended static checking for java,'' in \emph{Proceedings of the
  ACM SIGPLAN 2002 Conference on Programming language design and
  implementation}, 2002, pp. 234--245.

\bibitem{caballero2012undangle}
J.~Caballero, G.~Grieco, M.~Marron, and A.~Nappa, ``Undangle: early detection
  of dangling pointers in use-after-free and double-free vulnerabilities,'' in
  \emph{Proceedings of the 2012 International Symposium on Software Testing and
  Analysis}, 2012, pp. 133--143.

\bibitem{nethercote2003valgrind}
N.~Nethercote and J.~Seward, ``Valgrind: A program supervision framework,''
  \emph{Electronic notes in theoretical computer science}, vol.~89, no.~2, pp.
  44--66, 2003.

\bibitem{nethercote2007valgrind}
------, ``Valgrind: a framework for heavyweight dynamic binary
  instrumentation,'' \emph{ACM Sigplan notices}, vol.~42, no.~6, pp. 89--100,
  2007.

\bibitem{seward2005using}
J.~Seward and N.~Nethercote, ``Using valgrind to detect undefined value errors
  with bit-precision.'' in \emph{USENIX Annual Technical Conference, General
  Track}, 2005, pp. 17--30.

\bibitem{ven2006introduction}
K.~Ven, D.~V. Nuffel, and J.~Verelst, ``The introduction of openoffice. org in
  the brussels public administration,'' in \emph{IFIP International Conference
  on Open Source Systems}.\hskip 1em plus 0.5em minus 0.4em\relax Springer,
  2006, pp. 123--134.

\bibitem{hovemeyer2004finding}
D.~Hovemeyer and W.~Pugh, ``Finding bugs is easy,'' \emph{Acm sigplan notices},
  vol.~39, no.~12, pp. 92--106, 2004.

\bibitem{fu2021flowdist}
X.~Fu and H.~Cai,
  ``$\{$FlowDist$\}$:$\{$Multi-Staged$\}$$\{$Refinement-Based$\}$ dynamic
  information flow analysis for distributed software systems,'' in \emph{30th
  USENIX Security Symposium (USENIX Security 21)}, 2021, pp. 2093--2110.

\bibitem{blanchet2003static}
B.~Blanchet, P.~Cousot, R.~Cousot, J.~Feret, L.~Mauborgne, A.~Min{\'e},
  D.~Monniaux, and X.~Rival, ``A static analyzer for large safety-critical
  software,'' in \emph{Proceedings of the ACM SIGPLAN 2003 conference on
  Programming language design and implementation}, 2003, pp. 196--207.

\bibitem{jetley2008static}
R.~P. Jetley, P.~L. Jones, and P.~Anderson, ``Static analysis of medical device
  software using codesonar,'' in \emph{Proceedings of the 2008 workshop on
  Static analysis}, 2008, pp. 22--29.

\bibitem{serebryany2012addresssanitizer}
K.~Serebryany, D.~Bruening, A.~Potapenko, and D.~Vyukov,
  ``$\{$AddressSanitizer$\}$: A fast address sanity checker,'' in \emph{2012
  USENIX Annual Technical Conference (USENIX ATC 12)}, 2012, pp. 309--318.

\bibitem{harmim2019scalable}
D.~Harmim, V.~Marcin, and O.~Pavela, ``Scalable static analysis using facebook
  infer,'' \emph{Excel@ FIT19}, 2019.

\bibitem{kroening2016sound}
D.~Kroening, D.~Poetzl, P.~Schrammel, and B.~Wachter, ``Sound static deadlock
  analysis for c/pthreads,'' in \emph{Proceedings of the 31st IEEE/ACM
  International Conference on Automated Software Engineering}, 2016, pp.
  379--390.

\bibitem{simpson2013memsafe}
M.~S. Simpson and R.~K. Barua, ``Memsafe: ensuring the spatial and temporal
  memory safety of c at runtime,'' \emph{Software: Practice and Experience},
  vol.~43, no.~1, pp. 93--128, 2013.

\bibitem{lu2005bugbench}
S.~Lu, Z.~Li, F.~Qin, L.~Tan, P.~Zhou, and Y.~Zhou, ``Bugbench: Benchmarks for
  evaluating bug detection tools,'' in \emph{Workshop on the evaluation of
  software defect detection tools}, vol.~5.\hskip 1em plus 0.5em minus
  0.4em\relax Chicago, Illinois, 2005.

\bibitem{cadar2008klee}
C.~Cadar, D.~Dunbar, D.~R. Engler \emph{et~al.}, ``Klee: unassisted and
  automatic generation of high-coverage tests for complex systems programs.''
  in \emph{OSDI}, vol.~8, 2008, pp. 209--224.

\bibitem{li2020pca}
W.~Li, H.~Cai, Y.~Sui, and D.~Manz, ``Pca: memory leak detection using partial
  call-path analysis,'' in \emph{Proceedings of the 28th ACM Joint Meeting on
  European Software Engineering Conference and Symposium on the Foundations of
  Software Engineering}, 2020, pp. 1621--1625.

\bibitem{bruening2011practical}
D.~Bruening and Q.~Zhao, ``Practical memory checking with dr. memory,'' in
  \emph{International Symposium on Code Generation and Optimization (CGO
  2011)}.\hskip 1em plus 0.5em minus 0.4em\relax IEEE, 2011, pp. 213--223.

\bibitem{stepanov2015memorysanitizer}
E.~Stepanov and K.~Serebryany, ``Memorysanitizer: fast detector of
  uninitialized memory use in c++,'' in \emph{2015 IEEE/ACM International
  Symposium on Code Generation and Optimization (CGO)}.\hskip 1em plus 0.5em
  minus 0.4em\relax IEEE, 2015, pp. 46--55.

\bibitem{henning2000spec}
J.~L. Henning, ``Spec cpu2000: Measuring cpu performance in the new
  millennium,'' \emph{Computer}, vol.~33, no.~7, pp. 28--35, 2000.

\bibitem{henning2006spec}
------, ``Spec cpu2006 benchmark descriptions,'' \emph{ACM SIGARCH Computer
  Architecture News}, vol.~34, no.~4, pp. 1--17, 2006.

\bibitem{li2022polycruise}
W.~Li, J.~Ming, X.~Luo, and H.~Cai, ``$\{$PolyCruise$\}$: A
  $\{$Cross-Language$\}$ dynamic information flow analysis,'' in \emph{31st
  USENIX Security Symposium (USENIX Security 22)}, 2022, pp. 2513--2530.

\bibitem{wang2021bofsanitizer}
W.~Wang, M.~Fan, A.~Yu, and D.~Meng, ``Bofsanitizer: Efficient locator and
  detector for buffer overflow vulnerability,'' in \emph{2021 IEEE 23rd Int
  Conf on High Performance Computing \& Communications; 7th Int Conf on Data
  Science \& Systems; 19th Int Conf on Smart City; 7th Int Conf on
  Dependability in Sensor, Cloud \& Big Data Systems \& Application
  (HPCC/DSS/SmartCity/DependSys)}.\hskip 1em plus 0.5em minus 0.4em\relax IEEE,
  2021, pp. 1075--1083.

\bibitem{black2018juliet}
P.~E. Black and P.~E. Black, \emph{Juliet 1.3 test suite: Changes from
  1.2}.\hskip 1em plus 0.5em minus 0.4em\relax US Department of Commerce,
  National Institute of Standards and Technology, 2018.

\bibitem{salimi2022vulslicer}
S.~Salimi and M.~Kharrazi, ``Vulslicer: Vulnerability detection through code
  slicing,'' \emph{Journal of Systems and Software}, vol. 193, p. 111450, 2022.

\bibitem{xiao2020mvp}
Y.~Xiao, B.~Chen, C.~Yu, Z.~Xu, Z.~Yuan, F.~Li, B.~Liu, Y.~Liu, W.~Huo, W.~Zou
  \emph{et~al.}, ``$\{$MVP$\}$: Detecting vulnerabilities using
  $\{$Patch-Enhanced$\}$ vulnerability signatures,'' in \emph{29th USENIX
  Security Symposium (USENIX Security 20)}, 2020, pp. 1165--1182.

\bibitem{du2019leopard}
X.~Du, B.~Chen, Y.~Li, J.~Guo, Y.~Zhou, Y.~Liu, and Y.~Jiang, ``Leopard:
  Identifying vulnerable code for vulnerability assessment through program
  metrics,'' in \emph{2019 IEEE/ACM 41st International Conference on Software
  Engineering (ICSE)}.\hskip 1em plus 0.5em minus 0.4em\relax IEEE, 2019, pp.
  60--71.

\bibitem{booth2013national}
H.~Booth, D.~Rike, G.~A. Witte \emph{et~al.}, ``The national vulnerability
  database (nvd): Overview,'' 2013.

\bibitem{nong2022open}
Y.~Nong, R.~Sharma, A.~Hamou-Lhadj, X.~Luo, and H.~Cai, ``Open science in
  software engineering: A study on deep learning-based vulnerability
  detection,'' \emph{IEEE Transactions on Software Engineering}, 2022.

\bibitem{li2018vuldeepecker}
Z.~Li, D.~Zou, S.~Xu, X.~Ou, H.~Jin, S.~Wang, Z.~Deng, and Y.~Zhong,
  ``Vuldeepecker: A deep learning-based system for vulnerability detection,''
  \emph{arXiv preprint arXiv:1801.01681}, 2018.

\bibitem{black2017sard}
P.~E. Black \emph{et~al.}, ``Sard: A software assurance reference dataset,'' in
  \emph{Anonymous Cybersecurity Innovation Forum.()}, 2017.

\bibitem{nvd}
{National Institute of Standards and Technology (NIST)}, ``{National
  Vulnerability Database (NVD)},'' \url{https://nvd.nist.gov}, 2022.

\bibitem{li2021sysevr}
Z.~Li, D.~Zou, S.~Xu, H.~Jin, Y.~Zhu, and Z.~Chen, ``Sysevr: A framework for
  using deep learning to detect software vulnerabilities,'' \emph{IEEE
  Transactions on Dependable and Secure Computing}, 2021.

\bibitem{zou2019mu}
D.~Zou, S.~Wang, S.~Xu, Z.~Li, and H.~Jin, ``uvuldeepecker: A deep
  learning-based system for multiclass vulnerability detection,'' \emph{IEEE
  Transactions on Dependable and Secure Computing}, vol.~18, no.~5, pp.
  2224--2236, 2019.

\bibitem{zagane2020deep}
M.~Zagane, M.~K. Abdi, and M.~Alenezi, ``Deep learning for software
  vulnerabilities detection using code metrics,'' \emph{IEEE Access}, vol.~8,
  pp. 74\,562--74\,570, 2020.

\bibitem{wu2022vulcnn}
Y.~Wu, D.~Zou, S.~Dou, W.~Yang, D.~Xu, and H.~Jin, ``Vulcnn: An image-inspired
  scalable vulnerability detection system,'' 2022.

\bibitem{tang2022sevuldet}
Z.~Tang, Q.~Hu, Y.~Hu, W.~Kuang, and J.~Chen, ``Sevuldet: A semantics-enhanced
  learnable vulnerability detector,'' in \emph{2022 52nd Annual IEEE/IFIP
  International Conference on Dependable Systems and Networks (DSN)}.\hskip 1em
  plus 0.5em minus 0.4em\relax IEEE, 2022, pp. 150--162.

\bibitem{thapa2022transformer}
C.~Thapa, S.~I. Jang, M.~E. Ahmed, S.~Camtepe, J.~Pieprzyk, and S.~Nepal,
  ``Transformer-based language models for software vulnerability detection:
  Performance, model's security and platforms,'' \emph{arXiv preprint
  arXiv:2204.03214}, 2022.

\bibitem{cao2022mvd}
S.~Cao, X.~Sun, L.~Bo, R.~Wu, B.~Li, and C.~Tao, ``Mvd: Memory-related
  vulnerability detection based on flow-sensitive graph neural networks,''
  \emph{arXiv preprint arXiv:2203.02660}, 2022.

\bibitem{pinconschitenet}
E.~Pinconschi, S.~Reis, C.~Zhang, R.~Abreu, H.~Erdogmus, C.~S. Pasareanu, and
  L.~Jia, ``Tenet: A flexible framework for machine-learning-based
  vulnerability detection.''

\bibitem{chakraborty2021deep}
S.~Chakraborty, R.~Krishna, Y.~Ding, and B.~Ray, ``Deep learning based
  vulnerability detection: Are we there yet,'' \emph{IEEE Transactions on
  Software Engineering}, 2021.

\bibitem{russell2018automated}
R.~Russell, L.~Kim, L.~Hamilton, T.~Lazovich, J.~Harer, O.~Ozdemir,
  P.~Ellingwood, and M.~McConley, ``Automated vulnerability detection in source
  code using deep representation learning,'' in \emph{2018 17th IEEE
  international conference on machine learning and applications (ICMLA)}.\hskip
  1em plus 0.5em minus 0.4em\relax IEEE, 2018, pp. 757--762.

\bibitem{nong2020preliminary}
Y.~Nong and H.~Cai, ``A preliminary study on open-source memory vulnerability
  detectors,'' in \emph{2020 IEEE 27th International Conference on Software
  Analysis, Evolution and Reengineering (SANER)}.\hskip 1em plus 0.5em minus
  0.4em\relax IEEE, 2020, pp. 557--561.

\bibitem{nong2021evaluating}
Y.~Nong, H.~Cai, P.~Ye, L.~Li, and F.~Chen, ``Evaluating and comparing memory
  error vulnerability detectors,'' \emph{Information and Software Technology},
  vol. 137, p. 106614, 2021.

\bibitem{zhou2019devign}
Y.~Zhou, S.~Liu, J.~Siow, X.~Du, and Y.~Liu, ``Devign: Effective vulnerability
  identification by learning comprehensive program semantics via graph neural
  networks,'' \emph{arXiv preprint arXiv:1909.03496}, 2019.

\bibitem{lin2018cross}
G.~Lin, J.~Zhang, W.~Luo, L.~Pan, Y.~Xiang, O.~De~Vel, and P.~Montague,
  ``Cross-project transfer representation learning for vulnerable function
  discovery,'' \emph{IEEE Transactions on Industrial Informatics}, vol.~14,
  no.~7, pp. 3289--3297, 2018.

\bibitem{lin2019software}
G.~Lin, J.~Zhang, W.~Luo, L.~Pan, O.~De~Vel, P.~Montague, and Y.~Xiang,
  ``Software vulnerability discovery via learning multi-domain knowledge
  bases,'' \emph{IEEE Transactions on Dependable and Secure Computing}, 2019.

\bibitem{li2021vulnerability}
Y.~Li, S.~Wang, and T.~N. Nguyen, ``Vulnerability detection with fine-grained
  interpretations,'' in \emph{Proceedings of the 29th ACM Joint Meeting on
  European Software Engineering Conference and Symposium on the Foundations of
  Software Engineering}, 2021, pp. 292--303.

\bibitem{hin2022linevd}
D.~Hin, A.~Kan, H.~Chen, and M.~A. Babar, ``Linevd: Statement-level
  vulnerability detection using graph neural networks,'' \emph{arXiv preprint
  arXiv:2203.05181}, 2022.

\bibitem{zou2022mvulpreter}
D.~Zou, Y.~Hu, W.~Li, Y.~Wu, H.~Zhao, and H.~Jin, ``mvulpreter: A
  multi-granularity vulnerability detection system with interpretations,''
  \emph{IEEE Transactions on Dependable and Secure Computing}, no.~01, pp.
  1--12, 2022.

\bibitem{fu2022linevul}
M.~Fu and C.~Tantithamthavorn, ``Linevul: A transformer-based line-level
  vulnerability prediction,'' 2022.

\bibitem{cheng2022path}
X.~Cheng, G.~Zhang, H.~Wang, and Y.~Sui, ``Path-sensitive code embedding via
  contrastive learning for software vulnerability detection,'' in
  \emph{Proceedings of the 31st ACM SIGSOFT International Symposium on Software
  Testing and Analysis}, 2022, pp. 519--531.

\bibitem{fan2020ac}
J.~Fan, Y.~Li, S.~Wang, and T.~N. Nguyen, ``A c/c++ code vulnerability dataset
  with code changes and cve summaries,'' in \emph{Proceedings of the 17th
  International Conference on Mining Software Repositories}, 2020, pp.
  508--512.

\bibitem{li2020vuldeelocator}
Z.~Li, D.~Zou, S.~Xu, Z.~Chen, Y.~Zhu, and H.~Jin, ``Vuldeelocator: A deep
  learning-based fine-grained vulnerability detector,'' \emph{arXiv preprint
  arXiv:2001.02350}, 2020.

\bibitem{guo2019vulhunter}
N.~Guo, X.~Li, H.~Yin, and Y.~Gao, ``Vulhunter: An automated vulnerability
  detection system based on deep learning and bytecode,'' in
  \emph{International Conference on Information and Communications
  Security}.\hskip 1em plus 0.5em minus 0.4em\relax Springer, 2019, pp.
  199--218.

\bibitem{5552783}
N.~Antunes and M.~Vieira, ``Benchmarking vulnerability detection tools for web
  services,'' in \emph{2010 IEEE International Conference on Web Services},
  2010, pp. 203--210.

\bibitem{fonseca2007testing}
J.~Fonseca, M.~Vieira, and H.~Madeira, ``Testing and comparing web
  vulnerability scanning tools for sql injection and xss attacks,'' in
  \emph{13th Pacific Rim international symposium on dependable computing (PRDC
  2007)}.\hskip 1em plus 0.5em minus 0.4em\relax IEEE, 2007, pp. 365--372.

\bibitem{kang2022detecting}
H.~J. Kang, K.~L. Aw, and D.~Lo, ``Detecting false alarms from automatic static
  analysis tools: How far are we?'' \emph{arXiv preprint arXiv:2202.05982},
  2022.

\bibitem{shirey2000internet}
R.~Shirey, ``Internet security glossary,'' Tech. Rep., 2000.

\bibitem{antunes2010benchmarking}
N.~Antunes and M.~Vieira, ``Benchmarking vulnerability detection tools for web
  services,'' in \emph{2010 IEEE International Conference on Web
  Services}.\hskip 1em plus 0.5em minus 0.4em\relax IEEE, 2010, pp. 203--210.

\bibitem{li2019comparative}
Z.~Li, D.~Zou, J.~Tang, Z.~Zhang, M.~Sun, and H.~Jin, ``A comparative study of
  deep learning-based vulnerability detection system,'' \emph{IEEE Access},
  vol.~7, pp. 103\,184--103\,197, 2019.

\bibitem{fernandez2019case}
G.~C. Fern{\'a}ndez and S.~Xu, ``A case study on using deep learning for
  network intrusion detection,'' in \emph{MILCOM 2019-2019 IEEE Military
  Communications Conference (MILCOM)}.\hskip 1em plus 0.5em minus 0.4em\relax
  IEEE, 2019, pp. 1--6.

\bibitem{jabeen2022machine}
G.~Jabeen, S.~Rahim, W.~Afzal, D.~Khan, A.~A. Khan, Z.~Hussain, and T.~Bibi,
  ``Machine learning techniques for software vulnerability prediction: a
  comparative study,'' \emph{Applied Intelligence}, pp. 1--22, 2022.

\bibitem{zhang2023survey}
Q.~Zhang, C.~Fang, Y.~Ma, W.~Sun, and Z.~Chen, ``A survey of learning-based
  automated program repair,'' \emph{arXiv preprint arXiv:2301.03270}, 2023.

\bibitem{mazuera2021shallow}
A.~Mazuera-Rozo, A.~Mojica-Hanke, M.~Linares-V{\'a}squez, and G.~Bavota,
  ``Shallow or deep? an empirical study on detecting vulnerabilities using deep
  learning,'' in \emph{2021 IEEE/ACM 29th International Conference on Program
  Comprehension (ICPC)}.\hskip 1em plus 0.5em minus 0.4em\relax IEEE, 2021, pp.
  276--287.

\bibitem{steenhoek2022empirical}
B.~Steenhoek, M.~M. Rahman, R.~Jiles, and W.~Le, ``An empirical study of deep
  learning models for vulnerability detection,'' \emph{arXiv preprint
  arXiv:2212.08109}, 2022.

\bibitem{siow2022learning}
J.~K. Siow, S.~Liu, X.~Xie, G.~Meng, and Y.~Liu, ``Learning program semantics
  with code representations: An empirical study,'' in \emph{2022 IEEE
  International Conference on Software Analysis, Evolution and Reengineering
  (SANER)}.\hskip 1em plus 0.5em minus 0.4em\relax IEEE, 2022, pp. 554--565.

\bibitem{croft2023data}
R.~Croft, M.~A. Babar, and M.~Kholoosi, ``Data quality for software
  vulnerability datasets,'' \emph{arXiv preprint arXiv:2301.05456}, 2023.

\bibitem{liu2022investigating}
L.~Liu, Z.~Li, Y.~Wen, and P.~Chen, ``Investigating the impact of vulnerability
  datasets on deep learning-based vulnerability detectors,'' \emph{PeerJ
  Computer Science}, vol.~8, p. e975, 2022.

\bibitem{bhandari2021cvefixes}
G.~Bhandari, A.~Naseer, and L.~Moonen, ``Cvefixes: automated collection of
  vulnerabilities and their fixes from open-source software,'' in
  \emph{Proceedings of the 17th International Conference on Predictive Models
  and Data Analytics in Software Engineering}, 2021, pp. 30--39.

\bibitem{reis2021ground}
S.~Reis and R.~Abreu, ``A ground-truth dataset of real security patches,''
  \emph{arXiv preprint arXiv:2110.09635}, 2021.

\bibitem{xu2021tracer}
C.~Xu, B.~Chen, C.~Lu, K.~Huang, X.~Peng, and Y.~Liu, ``Tracer: Finding patches
  for open source software vulnerabilities,'' \emph{arXiv preprint
  arXiv:2112.02240}, 2021.

\bibitem{sawadogo2021early}
A.~D. Sawadogo, Q.~Guimard, T.~F. Bissyand{\'e}, A.~K. Kabor{\'e}, J.~Klein,
  and N.~Moha, ``Early detection of security-relevant bug reports using machine
  learning: How far are we?'' \emph{arXiv preprint arXiv:2112.10123}, 2021.

\bibitem{sawadogo2022sspcatcher}
A.~D. Sawadogo, T.~F. Bissyand{\'e}, N.~Moha, K.~Allix, J.~Klein, L.~Li, and
  Y.~Le~Traon, ``Sspcatcher: Learning to catch security patches,''
  \emph{Empirical Software Engineering}, vol.~27, no.~6, p. 151, 2022.

\bibitem{nguyen2022hermes}
G.~Nguyen-Truong, H.~J. Kang, D.~Lo, A.~Sharma, A.~E. Santosa, A.~Sharma, and
  M.~Y. Ang, ``Hermes: Using commit-issue linking to detect
  vulnerability-fixing commits,'' in \emph{2022 IEEE International Conference
  on Software Analysis, Evolution and Reengineering (SANER)}.\hskip 1em plus
  0.5em minus 0.4em\relax IEEE, 2022, pp. 51--62.

\bibitem{nguyen2022vulcurator}
T.~G. Nguyen, T.~Le-Cong, H.~J. Kang, X.-B.~D. Le, and D.~Lo, ``Vulcurator: a
  vulnerability-fixing commit detector,'' in \emph{Proceedings of the 30th ACM
  Joint European Software Engineering Conference and Symposium on the
  Foundations of Software Engineering}, 2022, pp. 1726--1730.

\bibitem{guo2022hyvuldect}
W.~Guo, Y.~Fang, C.~Huang, H.~Ou, C.~Lin, and Y.~Guo, ``Hyvuldect: A hybrid
  semantic vulnerability mining system based on graph neural network,''
  \emph{Computers \& Security}, p. 102823, 2022.

\bibitem{zheng2021d2a}
Y.~Zheng, S.~Pujar, B.~Lewis, L.~Buratti, E.~Epstein, B.~Yang, J.~Laredo,
  A.~Morari, and Z.~Su, ``D2a: A dataset built for ai-based vulnerability
  detection methods using differential analysis,'' in \emph{2021 IEEE/ACM 43rd
  International Conference on Software Engineering: Software Engineering in
  Practice (ICSE-SEIP)}.\hskip 1em plus 0.5em minus 0.4em\relax IEEE, 2021, pp.
  111--120.

\bibitem{woo2021v0finder}
S.~Woo, D.~Lee, S.~Park, H.~Lee, and S.~Dietrich, ``V0finder: Discovering the
  correct origin of publicly reported software vulnerabilities.'' in
  \emph{USENIX Security Symposium}, 2021, pp. 3041--3058.

\bibitem{li2022polyfax}
W.~Li, L.~Li, and H.~Cai, ``Polyfax: a toolkit for characterizing
  multi-language software,'' in \emph{Proceedings of the 30th ACM Joint
  European Software Engineering Conference and Symposium on the Foundations of
  Software Engineering}, 2022, pp. 1662--1666.

\bibitem{nong2022generating}
Y.~Nong, Y.~Ou, M.~Pradel, F.~Chen, and H.~Cai, ``Generating realistic
  vulnerabilities via neural code editing: An empirical study,'' in \emph{ACM
  Joint Meeting on European Software Engineering Conference and Symposium on
  the Foundations of Software Engineering (ESEC/FSE)}, 2022.

\bibitem{he2023controlling}
J.~He and M.~Vechev, ``Controlling large language models to generate secure and
  vulnerable code,'' \emph{arXiv preprint arXiv:2302.05319}, 2023.

\bibitem{nong2022vulgen}
Y.~Nong, Y.~Ou, M.~Pradel, F.~Chen, and H.~Cai, ``Vulgen: Realistic
  vulnerability generation via pattern mining and deep learning,'' \emph{in
  IEEE/ACM 45th International Conference on Software Engineering (ICSE)
  \url{https://www.software-lab.org/publications/icse2023_VulGen.pdf}}, 2023.

\end{thebibliography}
